\documentclass[twocolumn,showpacs,preprintnumbers,amsmath,amssymb]{revtex4}
\usepackage{amsfonts}
\usepackage{amsmath}
\usepackage{graphicx}
\usepackage{dcolumn}
\usepackage{bm}
\usepackage{overpic}
\usepackage{booktabs}

\begin{document}
\title{Identifying and Characterizing Nodes Important to Community Structure Using the Spectrum of the Graph}
\author{Yang Wang, Zengru Di, Ying Fan\footnote{yfan@bnu.edu.cn}}
\affiliation{Department of Systems Science, School of Management and Center for Complexity
Research, Beijing Normal University, Beijing 100875, China}
\date{\today}
\begin{abstract}
\textbf{Background:} Many complex systems can be represented as networks, and how a network breaks up into subnetworks or communities is of wide interest. However, the development of a method to detect nodes important to communities that is both fast and accurate is a very challenging and open problem.

\textbf{Methodology/Principal Findings}: In this manuscript, we introduce a new approach to characterize the node importance to communities. First, a centrality metric is proposed to measure the importance of network nodes to community structure using the spectrum of the adjacency matrix. We define the node importance to communities as the relative change in the eigenvalues of the network adjacency matrix upon their removal. Second, we also propose an index to distinguish two kinds of important nodes in communities, i.e., ``community core" and ``bridge".

\textbf{Conclusions/Significance}: Our indices are only relied on the spectrum of the graph matrix. They are applied in many artificial networks as well as many real-world networks. This new methodology gives us a basic approach to solve this challenging problem and provides a realistic result.
\end{abstract}

\pacs{89.75.Hc, 89.75.-k, 89.75.Fb}
\maketitle
\section{Introduction}

Networks, despite their simplicity, represent the interaction structure among components in a wide range of real complex systems, from social relationships among individuals, to interactions of proteins in biological systems, to the interdependence of function calls in large software projects. The network concept has been developed as an important tool for analyzing the relationship of structure and function for many complex systems in the last decades\cite{Rev.Mod.Phys.74,SIAM Rev.45,Science 286,Nature 393,Phys. Rep. 424}. Many real-world systems show the existence of structural modules that play significant and defined functional roles, such as friend groups in social networks, thematic clusters on the world wide web, functional groups in biochemical or neural networks\cite{PNAS99}. Exploring network communities is important for the reasons listed below\cite{Plos1}: 1) communities reveal the network at a coarse level, 2) communities provide a new aspect for understanding dynamic processes occurring in the network and 3) communities uncover relationships among the nodes that, although they can typically be attributed to the function of the system, are not apparent when inspecting the graph as a whole. As a result, it is not surprising that recent years have witnessed an explosion of research on community structure in graphs, and a huge number of methods or techniques have been designed\cite{PNAS99,Phys. Rev. E 74,Physics Reports486,Eur.Phys.J.B.38,Phys.Rev.E 72,Proc. Natl. Acad. 103,Phys. Rev. E 72,arXiv:1002.2007v1,arXiv:0902.3331v1,Rhys. Rev. E 77,arXiv:0907.3708}(see\cite{Physics Reports486} as a review).

It is believed that community structure is important to the function of a system\cite{Proc. Natl. Acad. Sci.100,Physica A 384,Europhys. Lett.72}. In many situations, it might be desirable to control the function of modular networks by adjusting the structure of communities. For example, in biological systems, one might like to identify the nodes that are key to communities and protect them or disrupt them, such as in the case of lung cancer\cite{Physica A 384}. In epidemic spreading, one would like to find the important nodes to understand the dynamic processes, which could yield an efficient method to immunize modular networks\cite{Europhys. Lett.72}. Such strategies would greatly benefit from a quantitative characterization of the node importance to community structure. Some important work related to this topic has been proposed. In 2006, Newman proposed a community-based metric called ``Community Centrality" to measure node importance to communities\cite{Phys. Rev. E 74}. His basic idea relies on the modularity function $Q$. Those vertices that contribute more to $Q$ are more important for the communities than those vertices that contribute less. Kovacs et al. also proposed an influence function to measure the node importance to communities\cite{PLOSONE}.

In fact, the important nodes can have distinct functions with respect to community structure. Some previous studies have also revealed such classifications. Guimera et al. have proposed a classification of the nodes based on their roles within communities, using their within-module degree and their participation coefficient\cite{Nature433}. They divided the hubs into three categories: provincial hubs, connector hubs and kinless hubs. Other approaches have also been suggested to discuss the connection between nodes and modularity in biological networks, by dividing hub nodes into two categories called ``party hubs" and ``date hubs"\cite{Nature430,PLoS Bioo,PLoS Bio}. When removed from the network, party and date hubs have strikingly distinct effects on the overall topology of the network. Recently, Kovacs et al. proposed an interesting approach. They introduced an integrative method family to detect the key nodes, overlapping communities and ``date" and ``party" hubs\cite{PLOSONE}. In a very recent work, the authors mentioned that modular networks naturally allow the formation of clusters, and hubs connecting the modules would enhance the integration of the whole network, such as in the case of neuron networks\cite{PRE82}. As a result, it is intuitive that nodes that are important to communities can be divided into ``community cores" and ``bridges". However, there is one problem. Before using the participation coefficient and the influence function to distinguish these two kinds of vertices, the exact communities of the network must first be given. In contrast, it is interesting to characterize node importance to communities before the division of the network.

It is understood that the adjacency matrix contains all the information of the network. Developing methods based only on the adjacency matrix of the network to detect important nodes to communities and then distinguish them as either ``community core" or ``bridge" is an interesting and important problem in network research. In this manuscript, based only on the adjacency matrix of the network, we try to access the fundamental questions: how to evaluate the node importance to communities and how to distinguish different kinds of important nodes? It is implied that in many cases the spectrum of the adjacency matrix gives an indication of the community structure in the network\cite{PRE80}. If the network has $c$ strong communities, the $c$ largest eigenvalues of the adjacency matrix are significantly larger than the magnitudes of all the other eigenvalues. These large eigenvalues are key quantities to the community structure. For this reason, we suggest a basic approach to solve the above open problem using the spectrum of the graph. We define the importance of nodes to communities as the relative change in the $c$ largest eigenvalues of the network adjacency matrix upon their removal. Furthermore, using the eigenvectors of the graph Laplacian, we divide the important nodes into community cores and bridges. We apply our method to many networks, including artificial networks and real-world networks. This new methodology gives us a basic approach to solve this challenging problem and provides a realistic result.

The organization of this paper is as follows. In section \uppercase\expandafter{\romannumeral2}, the centrality metric identifying the important nodes to communities is proposed using the spectrum of the adjacency matrix. An index to distinguish the two kinds of important nodes using the corresponding eigenvector of the graph Laplacian is introduced in section \uppercase\expandafter{\romannumeral3}. In section \uppercase\expandafter{\romannumeral4}, our method is applied to artificial networks and some real-world networks, and we obtain some interesting results. In section \uppercase\expandafter{\romannumeral5}, we extend our method into weighted networks. Finally, concluding remarks are presented in section \uppercase\expandafter{\romannumeral5}.

\section{Centrality Metric Based on the Spectrum of the Adjacency Matrix}

We consider a binary network $G=(V,E)$ with $N$ nodes. The adjacency matrix $A$ is the matrix with elements $A_{ij} = 1$ if there is an edge joining vertices $i$ and $j$, otherwise $0$. We denote each eigenvalue of $A$ by $\lambda$ and the corresponding eigenvector by $\textit{\textbf{v}}$, such that $A \textit{\textbf{v}}= \lambda \textit{\textbf{v}}$. The eigenvector is orthogonal and normalized. The eigenvalues are ordered by decreasing magnitude: $\lambda _1 \ge \lambda _2  \ge  \cdots  \ge \lambda _n$. It is easy to show that $A$ is symmetric and the eigenvalues of $A$ are real. Consider the case of networks that have $c$ communities. It is implied that when these communities are disconnected, each one has its own largest eigenvalues. With proper labeling of the nodes, the matrix $A$ will have a block matrix structure with $c \times c$ blocks. Blocks on the diagonal correspond to the adjacency matrices of the individual communities, while the off-diagonal blocks correspond to the edges between communities; in other words, we can consider them as a perturbation. Therefore, $A$ can be written as
\begin{equation}
    A=A_0+\delta A,
\end{equation}
where $A_0$ is a matrix whose diagonal block elements are the diagonal block elements of $A$ and whose off-diagonal block elements are zeros, while $\delta A$ is a matrix with zeros on its diagonal blocks and with the off-diagonal blocks of $A$ as its off-diagonal block elements. Chauhan et al.\cite{PRE80} have proved that if the perturbation strength is small, the largest eigenvalues of disconnected communities are perturbed more weakly than the perturbation applied. The spectrum of the adjacency matrix of a network gives a clear indication of the number of communities in the network. If the network has $c$ strong communities, the $c$ largest eigenvalues are well separated from others. These eigenvalues are key quantities to the community structure.

For this reason, we define the importance of node $k$ to communities as the relative change in the $c$ largest eigenvalues of the network adjacency matrix upon its removal:
\begin{equation}
I_k  = - \sum\limits_{i = 1}^c {\frac{{\Delta \lambda _i }}{{\lambda _i }}},
\end{equation}
where $c$ is the number of communities. To avoid the computational cost, we use perturbation theory to provide approximations of $I_k$ in terms of the corresponding eigenvector $\textit{\textbf{v}}$. Let us denote the matrix before the removal of the node by $A$ and the matrix after the removal by $A + \Delta A$; the eigenvalue of this matrix is $\lambda  + \Delta \lambda$, and the corresponding eigenvector is $\textit{\textbf{v}} + \Delta \textit{\textbf{v}}$. For large matrices, it is reasonable to assume that the removal of a node has a small effect on the whole matrix and the spectral properties of the network, so that $\Delta A$ and $\Delta \lambda$ are small. We obtain
\begin{equation}
(A + \Delta A)(\textit{\textbf{v}} + \Delta \textit{\textbf{v}}) = (\lambda  + \Delta \lambda )(\textit{\textbf{v}} + \Delta \textit{\textbf{v}}).
\end{equation}

The effect on the adjacency matrix $A$ of removing node $k$ is given by $(\Delta A)_{ij}  =  - A_{ij} (\delta _{ik}  + \delta _{jk} )$. We cannot assume that the $\Delta \textit{\textbf{v}}$ is small because $\Delta v_k = -v_k$, so we set $\Delta \textit{\textbf{v}} = \delta \textit{\textbf{v}} - v_k \widehat{e}_k$ where $\delta \textit{\textbf{v}}$ is small and $\widehat{\textit{\textbf{e}}}$ is the unit vector for the $k$ component. Left multiplying (3) by $\textit{\textbf{v}}^T$ and neglecting second order terms $\textit{\textbf{v}}^T \Delta A\delta \textit{\textbf{v}}$ and $\textit{\textbf{v}}^T \Delta \lambda \delta \textit{\textbf{v}}$, we obtain
\begin{equation}
\Delta \lambda  = \frac{{\textit{\textbf{v}}^T \Delta A\textit{\textbf{v}} - \textit{\textbf{v}}^T v_k \Delta A\widehat{e}_k }}{{\textit{\textbf{v}}^T \textit{\textbf{v}} - v_k^2 }}.
\end{equation}

For a large network ($N \gg 1$), we know that $\textit{\textbf{v}}^T \textit{\textbf{v}}\gg v_k^2$; therefore, we can write
\begin{equation}
\Delta \lambda  \approx \frac{{\textit{\textbf{v}}^T \Delta A\textit{\textbf{v}} - \textit{\textbf{v}} ^T \textit{\textbf{v}}_k \Delta A\widehat{\textit{\textbf{e}}}_k }}{{\textit{\textbf{v}}^T \textit{\textbf{v}}}}
\end{equation}
Because $(\Delta A)_{ij}  =  - A_{ij} (\delta _{ik}  + \delta _{jk} )$, we obtain
\begin{equation}
\textit{\textbf{v}}^T \Delta A\textit{\textbf{v}} =  - 2\lambda v_k^2,
\textit{\textbf{v}}^T v_k \Delta A\widehat{e}_k  =  - \lambda v_k^2.
\end{equation}
Finally, the importance of node $k$ to the community structure is obtained by
\begin{equation}
I_k  = -\sum\limits_{i = 1}^c {\frac{{\Delta \lambda _i }}{{\lambda _i }}}\approx
\sum\limits_{i = 1}^c {\frac{{v_{ik}^2 }}{{\textit{\textbf{v}}_i^T \textit{\textbf{v}}_i }}},
\end{equation}
where $c$ is the number of communities, $v_{ik}$ is the $\textit{k}$th element of $\textit{\textbf{v}}_i$ and $I_k$ lies in the interval $[0,1]$. If $I_k$ is large, node $\textit{k}$ is important to the community structure; otherwise, $k$ is on the periphery of the community.

Using this metric $I$, we can quantify the node importance to the community structure. If the node is important to the community structure, when we remove it from the network, the relative changes of the $c$ largest eigenvalues are large; otherwise, the changes are small. Before applying $I$, the value of $c$ needs to be determined. The determination of the number of communities is an important but challenging question in community analysis. Here we use the method proposed by Ref.\cite{PRE80}. This method is based on the properties of the spectrum of the graph and is independent of the partition algorithms, so our metric is quite convenient to use.

\section {distinguish two kinds of important nodes}

As mentioned above, there are two kinds of nodes that are important to communities. One is the ``community core", and the other is the ``bridge" between communities. Each will affect communities deeply upon its removal. When we remove the ``community core", the community structure in the network will become fuzzy, while the community structure will become clear when we remove the ``bridge". See Fig. 1 for an example. Vertices 1 and 8 are the ``community cores", and they organize their respective communities. Meanwhile, node 15 is the ``bridge" between the two communities. The ``community core" is the leader in the community, and it can organize the function of each community. In contrast, the ``bridge" connects the modules and can enhance the integration of the whole network. It is believed that a combination of both segregation and integration, as in neural systems, is crucial\cite{PRE82}. It is clear that effectively disconnected and fully non-synchronous regions cannot allow collective or integrative action of the elements. Similarly, a fully synchronized regime does not allow separated or segregated performance of the elements. Therefore, both situations are biologically unrealistic, as can be seen from the existence of related conditions, such as epileptic seizures (collective phenomena) and Parkinson's disease (segregated phenomena)\cite{Neuro}. For this reason, both the ``community core" and the ``bridge" are important to communities, but they play different roles. The metric we proposed in Section\uppercase\expandafter{\romannumeral2} can determine the nodes that are important to communities, but now a method to distinguish these two kinds of important nodes is needed.

In agreement with earlier findings\cite{PLOSONE,Nature430,PLoS Bioo,PLoS Bio}, we assumed that bridge nodes should have more inter-modular positions than community cores. The existence of bridge nodes often leads to some inter-modular edges. Given a graph, the simplest and most direct way to construct a partition of the graph is to solve the mincut problem (minimize the number of edges between communities $R$)\cite{CMJ23}. In practice, however, this method often does not lead to satisfactory partitions. The problem is that, in many cases, the solution of mincut simply separates one individual vertex from the rest of the graph. Of course, this is not what we want to achieve in clustering, as clusters should be reasonably large groups of points. Due to this shortcoming in the mincut problem, one common objective function to encode the desired information is RatioCut\cite{Ratiocut}:
\begin{equation}
RatioCut(C_1 , \cdots C_c ) \buildrel\textstyle.\over= \sum\limits_{i = 1}^c {\frac{{R(C_i ,\bar C_i )}}{{|C_i |}}},
\end{equation}
where $|C_i|$ is the size of community $C_i$. If the sizes of the communities are almost the same, the RatioCut problem reduces to the mincut problem.

\subsection{The Condition of $c = 2$}

If the network is divided into only two communities ($c = 2$), we define an index vector $\textit{\textbf{s}}$ with $N$ elements:
\begin{equation}
s_i  = \left\{ \begin{array}{l}
 \sqrt {{{|\bar C|} \mathord{\left/
 {\vphantom {{|\bar C|} {|C|}}} \right.
 \kern-\nulldelimiterspace} {|C|}}}\quad \quad {\rm{  if\quad vertex }}\quad i  \in C, \\
  - \sqrt {{{|C|} \mathord{\left/
 {\vphantom {{|C|} {|\bar C|}}} \right.
 \kern-\nulldelimiterspace} {|\bar C|}}} \quad {\rm{  if\quad vertex }}\quad i \in \bar C. \\
 \end{array} \right.
\end{equation}
Then the RatioCut function is obtained as follows\cite{Tutorial}:

\begin{equation}
RatioCut(C,\bar C) = \frac{1}{{|V|}}\textit{\textbf{s}}^T \textit{\textbf{L}}\textit{\textbf{s}},
\end{equation}
where $|V|$ is the number of vertices in the network and $\textit{\textbf{L}}$ is the graph Laplacian. $\textit{\textbf{L}}$ is defined as $L_{ij}=-A_{ij}$ for $i \ne j$ and $L_{ii}=k_i$, where $k_i$ is the degree of node $i$. We also have two constraints on $\textit{\textbf{s}}$: $\sum\limits_{i = 1}^n {s_i }  = 0$ and $\sum\limits_{i = 1}^n {s_i^2 }  = n$. Here the partition problem is equal to the problem
\begin{equation}
\min {\rm{ }}\textit{\textbf{s}}^T \textit{\textbf{L}}\textit{\textbf{s}};\ {\rm{ subject\ to }}\ \sum\limits_{i = 1}^n {s_i }  = 0,\sum\limits_{i = 1}^n {s_i^2 }  = n.
\end{equation}
If the components of the vector $\textit{\textbf{s}}$ are allowed to take arbitrary values, it can be seen immediately that the solution of this problem is given by the vector $\textit{\textbf{s}}$ that is the eigenvector corresponding to the second-smallest eigenvalue of $\textit{\textbf{L}}$, denoted by $\textit{\textbf{u}}_2$. So we can approximate a minimizer of RatioCut by the second eigenvector of $\textit{\textbf{L}}$. Unfortunately, the components of $\textit{\textbf{s}}$ are only allowed to take two particular values.

Thus, the simplest solution is achieved by assigning vertices to one of the groups according to the sign of the eigenvector $\textit{\textbf{u}}_2$. In other words, we assign vertices as follows: if $\textit{\textbf{u}}_2^i > 0$, we assign vertex $i$ to community $C$; otherwise, we assign it to $\bar C$. Assignation priority begins with the most positive and the most negative; the node with the most positive magnitude is first to be assigned to $C$, then the second and so on, while the node with the most negative magnitude is similarly the first to be assigned to $\bar C$.
If a node's corresponding element is close to zero, it may have nearly equal membership in both communities, and we can assign it to both communities. In conclusion, if the network is divided into only two communities, we can use this method to characterize which are the ``community cores" and which are the ``bridge" between communities. If node $i$ is a ``community core", $|\textit{\textbf{u}}_2^i|$ is relatively large; otherwise, $|\textit{\textbf{u}}_2^i|$ is near zero.

\subsection{The Condition of $c > 2$}

Consider the division of a network into $c$ nonoverlapping communities, where $c$ is the number of communities. We define an $n \times c$-index matrix $\textit{\textbf{S}}$ with one column for each community, $\textit{\textbf{S}} = (\textit{\textbf{s}}_1 |\textit{\textbf{s}}_2 | \cdots |\textit{\textbf{s}}_c )$, by
\begin{equation}
s_{i,j}  = \left\{ \begin{array}{l}
 {{\rm{1}} \mathord{\left/
 {\vphantom {{\rm{1}} {\sqrt {|C_j |} }}} \right.
 \kern-\nulldelimiterspace} {\sqrt {|C_j |} }} \quad{\rm{ if \quad vertex }}\quad i  \in C_j, \\
 {\rm{0 \quad  otherwise}}. \\
 \end{array} \right.
\end{equation}
Following the previous section, we obtain
\begin{equation}
RatioCut = Tr(\textit{\textbf{S}}^T \textit{\textbf{L}}\textit{\textbf{S}}),
\end{equation}
where $Tr$ is the trace of a matrix and $\textit{\textbf{S}}^T$ is the transpose matrix of $\textit{\textbf{S}}$. $\textit{\textbf{L}}$ is a semi-positive and symmetric matrix. We can write $\textit{\textbf{L}}=\textit{\textbf{U}}\textit{\textbf{D}}\textit{\textbf{U}}^T$, where $\textit{\textbf{U}}$ is the eigenvector of $\textit{\textbf{L}}$, $\textit{\textbf{U}} = (\textit{\textbf{u}}_1 |\textit{\textbf{u}}_2 | \cdots |\textit{\textbf{u}}_n )$ and $\textit{\textbf{D}}$ is the diagonal matrix of eigenvalues $D_{ii} = \beta_i$. We therefore obtain
\begin{equation}
RatioCut = \sum\limits_{j = 1}^n {\sum\limits_{k = 1}^c {\beta _j (u_j^T s_k )^2 } }.
\end{equation}
It can also be written as
\begin{equation}
RatioCut = \sum\limits_{k = 1}^c {\sum\limits_{j = 1}^n {\beta _j [\sum\limits_{i = 1}^n {U_{ij} S_{ik} } ]^2 } }.
\end{equation}
Now we define the vertex vector of $i$ as $r_i$, and let
\begin{equation}
[r_i ]_j  = U_{ij}.
\end{equation}
If the network has almost equal-sized communities, then equation (15) can be written as
\begin{equation}
RatioCut \approx \frac{{\sum\limits_{k = 1}^c {\sum\limits_{j = 1}^n {\beta _j [\sum\limits_{i \in G_k } {[r_i ]_j } ]^2 } } }}{{|C|}},
\end{equation}
where $G_k$ is the set of vertices belonging to community $k$ and $|C|$ is the community size.

Minimizing the RatioCut can be equated with the task of choosing the nonnegative quantities so as to place as much of the weight as possible in the terms corresponding to the low eigenvalues and as little as possible in the terms corresponding to the high eigenvalues. This equates to the following maximization problem:
\begin{equation}
Max\ \sum\limits_{k = 1}^c {\sum\limits_{j = 1}^p {\beta _j [\sum\limits_{i \in G_k } {[r_i ]_j } ]^2 } },
\end{equation}
where $p$ is a parameter. We could choose $p=c$ if the community structure was clear. To this end, we propose an easy way to distinguish two kinds of important nodes using the theory of the graph Laplacian. If the community structure is quite clear, we focus on the \textit{vertex vector magnitude} $|r_i |$ in the first $p$ terms, denoted by the $w$-score:
\begin{equation}
w_i = \sqrt {\sum\limits_{j = 1}^p {[r_i ]_j^2 } }.
\end{equation}
If the $w$-score of a given vertex is close to zero, we believe that this vertex has nearly equal membership in more than one community, and it is likely to be the ``bridge" of these communities. This discrimination process equates to the ``fuzzy" division of the network into communities. In many cases, this type of fuzzy division could result in a more accurate picture of real-world networks.

\section{Results}

Now we test the validity of our indices introduced in section \uppercase\expandafter{\romannumeral2} and section \uppercase\expandafter{\romannumeral3} in various artificial networks and real-world networks.

\subsection{Artificial Networks}

First, we consider a sketch composed of 15 nodes (see Fig. 1) forming two communities. It is intuitive that vertices 1, 8 and 15 are important to the community structure in this sketch. Vertices 1 and 8 are the so-called ``community cores", and they form both the communities. Vertex 15 is the ``bridge" between communities, and it connects these two communities. As we discussed before, removing vertex 1 or 8 will make the community structure fuzzy, and removing vertex 15 will make it clear.
\begin{figure}
\center\includegraphics[width=0.5\textwidth]{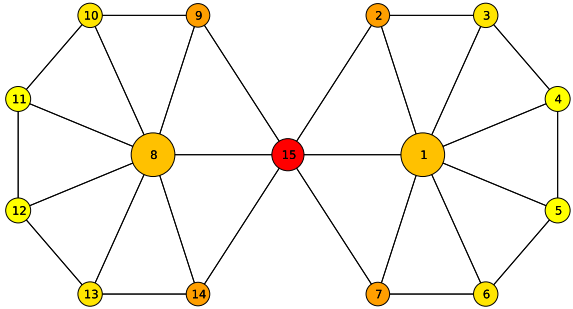}
\caption{Sketch of a network composed of 15 nodes. The diameter of one vertex is proportional to the centrality metric $I$. Moreover, the color of one vertex is related to the index $w$-score. Red vertices behave like ``overlapping" nodes or ``bridges" between communities, and yellow vertices often lie inside their own communities.}
\end{figure}
Here we use the index $H$ proposed by Hu et al.\cite{arXiv:1002.2007v1} to measure the significance of communities:
\begin{equation}\label{R}
H = \frac{n}{{\bar k\sum\limits_{j = c + 1}^n {\frac{1}{{|\overline \beta   - \beta _j |}}} }},
\end{equation}
where $\beta$ is the eigenvalue of the graph Laplacian,
$\overline \beta $
is the average value of $\beta_2$ through $\beta_c$, $\bar k$ is the average degree of the network and $n$ is the number of vertices in the network. In networks with strong communities (many links are within communities with very sparse connections outside), $H$ is always large. Here we focus on the change of $H$ due to the removal of vertices, denoted by $\Delta H$. We also use the centrality metric proposed by Newman\cite{Phys. Rev. E 74}, which we denote here by $M$. The results are shown in Tab. 1. Through $\Delta H$, it is implied that vertices 1 and 8 are more important than other vertices because the magnitude of $\Delta H$ is relatively larger than others. Moreover, their removal makes the communities fuzzy, while vertex 15 acts like a "bridge" between the communities, and its removal makes the communities clear. We can see that our centrality metric performs quite well; it can identify not only the ``community cores", but also the ``bridge" between communities. $M$ can also identify the ``community cores", but it has some problems. One issue is that its values tend to span a rather small dynamic range from largest to smallest. Moreover, in some cases (such as this sketch), $M$ cannot recognize important vertices among communities. In calculating the index $H$, we need to go through every vertex in the network, incurring significant computational cost. In contrast, our method provides a more efficient way, requiring less computational cost, and yields the correct answer.

\begin{table}[htbp]
\begin{center}
\caption{Centrality metrics of the example sketched in Fig. 1.}

\begin{tabular}{p{2cm} p{1cm} p{1cm} p{1cm} p{1.5cm}}
\hline
\hline
Vertex Label & $I$ & $M$ & $\Delta H$ & $w$-score  \\
\hline
1 & 0.32  &0.758 & -0.145 & 0.2405\\
8 & 0.32  &0.758 & -0.145 & 0.2405\\
15& 0.173 & 0.69 & 0.116 &  0.00\\
2,7,9,14 & 0.09  &0.704 &0.04&0.198\\
3,6,10,13& 0.1 & 0.7535 &-0.021 &0.285\\
4,5,11,12& 0.105&0.7327  &-0.054 & 0.3175\\
\hline
\hline
\end{tabular}
\vspace*{0.0cm}
\end{center}
\end{table}

Here we use the classical \textbf{GN benchmark} presented by Girvens and Newman to test the measurements\cite{Proc. Natl. Acad. 103}. Each network has $N = 128$ nodes that are divided into four communities (c = 4) with 32 nodes each. Edges between two nodes are introduced with different probabilities, which depend on whether the two nodes belong to the same community or not. Each node has $<k_{in}>$ links on average with its fellows in the same community and
$<k_{out}>$ links with the other communities, and we impose $<k_{in}>+<k_{out}> = 16$. The communities become fuzzier and thus more difficult to identify as $k_{out}$ increases. Because the GN benchmark is a homogenous network, there should not be any nodes that are important to the community structure. To check whether our conjecture is correct or not, we let $<k_{in}> = 12$ so that the community structure is quite clear and average the result for the GN benchmark over 100 configurations of networks. From the result, about 120 nodes' importances lie in the interval $[0.03,0.04]$, while others lie in the interval [0.02,0.03]. The mean value of $I$ is 0.0312, and the standard deviation is 0.0014. It can be concluded that, in the GN benchmark, there are no nodes that are important to the community structure.

We may also test the method on the more challenging \textbf{LFR benchmark} presented by Lancichinetti et al.\cite{Phys. Rev. E 78}. In the LFR benchmark, the degree distribution obeys a power-law distribution $p(k) \propto k^{ - \alpha }$, and the sizes of the communities are also taken from a power-law distribution with an exponent $\gamma$.
Moreover, each node shares a fraction $1-\mu$ of its links with other nodes of its own community and a fraction $\mu$ with others in the rest of the network. The community structure can be adjusted by the mixing parameter $\mu$. Without loss of generality, we let $\alpha = 2.5, \gamma = 1.0, \mu = 0.25$ and the size of the network $N = 1000$. Our numerical results in the LFR benchmark are shown in Fig. 2. In this case, there is no ``bridge" between communities because $\mu = 0.25$. We may also calculate the $w$-score, of which the mean value is 0.1736 and the standard deviation is 0.0292. Moreover, the centrality metric is positively correlated with node degree ($r^2 = 0.7329$), but some vertices have quite high centrality while having relatively low degree, and thus the correlation index is not very high.

\begin{figure}
\center\includegraphics[width=0.5\textwidth]{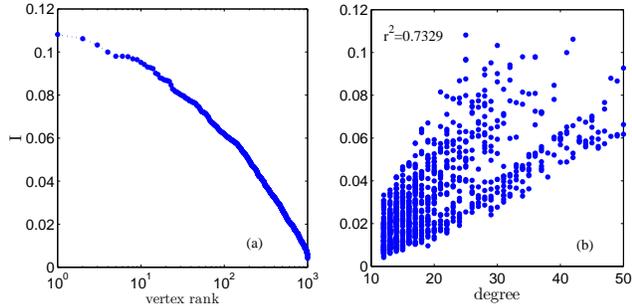}
\caption{(a) The Zipf plot of the nodes' centrality to communities. (b) The centrality metric we propose is correlated with node degree. The parameters in the LFR benchmark are as follows: $\alpha = 2.5, \gamma = 1.0, \mu = 0.25$ and the size of the network $N = 1000$.}
\end{figure}

\subsection{Real-world Networks}

We apply our method to some real-world networks, such as the Zachary club network\cite{JAR33}, the word association network\cite{SOUTH}, the scientific collaboration network\cite{website}, and the C. elegans neural network\cite{TRS}.

First, we consider a famous example of a social network, the \textbf{Zachary's karate club network}. This network represents the pattern of friendships among members of a karate club at a North American university. It contains 34 vertices, and the links between vertices are the friendships between people. The nodes labeled as 1 and 34 correspond to the club instructor and the administrator, respectively. They had a conflict which resulted in the breakup of the club. Most other nodes have a relationship with node 1, node 34, or both. In this network, $c = 2$. The numerical results are shown in Fig. 3 and Fig. 4. In Fig. 3(a), we can see that nodes 1 and 34 are the most important nodes in the communities. Our method to distinguish important nodes are shown in Fig. 3(b). From the result, we can see that nodes 1 and 34 are the so-called ``community cores", and they have many connections in their own communities. Furthermore, we compare our method with Newman's. This result is also shown in Fig. 3(a), and the two metrics are normalized by
\begin{equation}
x_{nor}=\frac{{x -  < x > }}{{\sigma _x }},
\end{equation}
where $< x >$ is the average value of each index and $\sigma_x$ is the standard deviation of each index. It is implied that these two methods have some differences. In our method, nodes 1 and 34 are absolutely more important than other nodes, while in Newman's method, nodes 2 and 33 are also quite important, even more than node 1. In this network, the modularity function $Q$ reaches its maximum value when the network is divided into 4 communities; this fact may be the cause of the differences between the results of these two methods. The visualization of the karate network with our two measurements is sketched in Fig. 4. The diameter of each vertex is proportional to the centrality metric $I$. A large diameter indicates an important vertex. Additionally, the color of each vertex is related to the index $w$-score. Red vertices behave like ``overlapping" nodes or ``bridges" between communities, and yellow vertices often lie inside their own communities.

\begin{figure}
\center\includegraphics[width=0.5\textwidth]{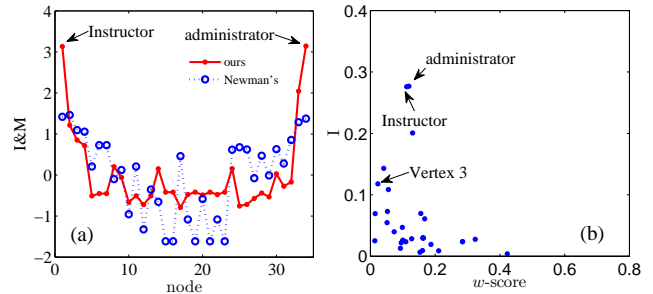}
\caption{It is shown that our method works quite well in the Zachary's karate club network. Nodes 1 and 34 are the instructor and the administrator, respectively. In Fig. 3(a), we can see that these two nodes are more important to the community structure than other nodes. We also compare our method with Newman's and find that the two methods exhibit some differences. In Fig. 3(b), we shown that nodes 1 and 34 are the so-called ``community cores".}
\end{figure}

\begin{figure}
\begin{center}
  \includegraphics[width=0.5\textwidth]{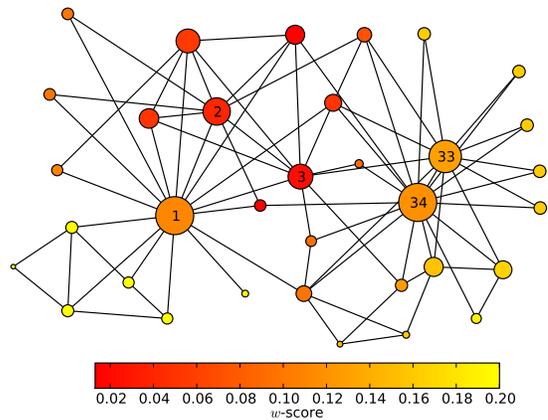}
  \caption{The Zachary's karate club network, which is composed of 34 vertices. Vertex diameters indicate the community centrality $I$. The color of each vertex is proportional to the index $w$-score.}
\end{center}
\end{figure}

Second, we analyze the \textbf{word association network} starting from the word ``Bright''. This network was built on the University of South Florida Free Association Norms\cite{SOUTH}. An edge between words A and B indicates that some people associate the word B to the word A. The graph displays four communities, corresponding to the categories \textit{Intelligence, Astronomy, Light, Colors}. The word \textit{Bright} is related to all of them by construction. We applied our method to this network, and the results are shown in Fig. 5. From the results, we can observe that our method considers \textit{Bright, Sun, Smart, Moon} as important nodes to the community structure. It may be inferred from the result that \textit{Moon} and \textit{Smart} are the ``community cores", while \textit{Bright} and \textit{Sun} are the ``bridges" between communities. Indeed, our metric yields the correct answer. For example, \textit{Smart} is the core of the community \textit{Intelligence}, while \textit{Moon} is the core of the community \textit{Astronomy}. Meanwhile, the $w$-score of node \textit{Bright} is 0.08, which is close to zero. We would therefore conclude that it is a ``bridge" between communities, and \textit{Bright} is in fact the ``bridge" among these four communities, as the network was originally derived from it.

Moreover, we may investigate the effect of node removal on the modularity function $Q$. ``Community cores" and ``bridges" have different effects on community structure. When a ``community core" is removed, the communities become clear. For example, the removal of the node ``bright" makes the modularity function $Q$ increase by 0.03, which is the largest increase caused by the removal of any single node, while the removal of node ``Moon" causes $Q$ to decrease by 0.015. These results are averaged over 20 trials. We can see from our results that important nodes (i.e., nodes with large $I$) affect the communities considerably. For example, the removal of the node ``Smart" decreases $Q$ by 0.0152, while the removal of the node ``Gifted", which seems to be a peripheral node, decreases $Q$ by only 0.0048.

\begin{figure}
\center\includegraphics[width=0.5\textwidth]{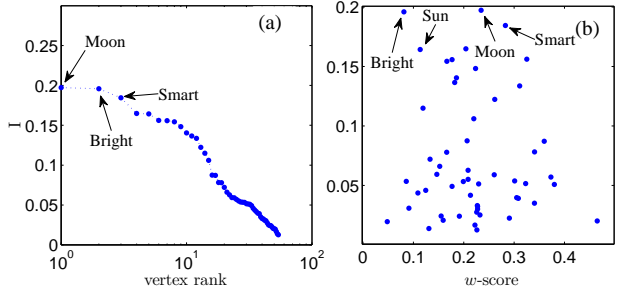}
\caption{Index $I$ and $\omega$-score for the nodes of the word association network. The node importance versus vertex rank is shown in (a). In (b), we distinguish ``community cores" and ``bridges" using the index $w$-score. }
\end{figure}

\begin{figure*}
\center\includegraphics[width=0.75\textwidth]{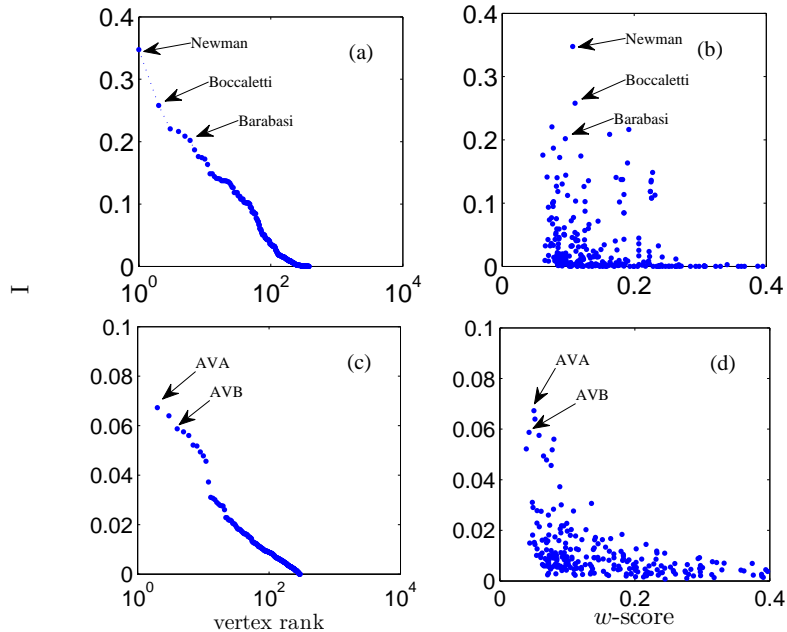}
\caption{The centrality metric $I$ and $w$-score for the scientist collaboration network (a,b). The centrality metric $I$ and $w$-score are also calculated in the C. elegans neural network (c,d).}
\end{figure*}

We may also apply our method to social networks, such as the \textbf{scientist collaboration network}\cite{website}, and neural networks, such as the \textbf{C. elegans neural network}\cite{TRS}. We analyzed the largest connected component of each network. The scientist collaboration network represents scientists whose research centers on the properties of networks of one kind or another. There are 379 vertices, representing scientists who are divided into 12 communities. Edges are placed between scientists who have published at least one paper together. The neural network of C. elegans contains 302 neurons and 2,359 links. This network is divided into 3 communities, with each node representing a neuron and each link representing a synaptic connection between neurons. Here we consider the C. elegans neural network to be undirected. The results are shown in Fig. 6.

In the scientist collaboration network, our centrality metric $I$ identifies ``group leaders", such as M. Newman, S. Boccaletti, and A. Barabasi. Their $w$-scores are not very large because they often have some collaboration between scientists outside their own communities. We can also find so-called ``community cores" based on our method, such as R. Sole, and ``bridge" vertices among some communities, such as B. Kahng. As we know, the C. elegans neural networks are composed of sensory neurons, interneurons and motor neurons. The neurons with high centrality metrics often have the most important functions, and all of them are interneurons, such as $AVA$, $AVB$, $AVD$, and $AVE$. These classes, which synapse onto motor neurons in the ventral cord, are among the most prominent neurons in the whole nervous system. They generally have larger-diameter processes than other neurons and have many synaptic connections\cite{TRS,JN}. As a result, they have larger $I$ than other vertices, while the typical $w$-score in these classes is quite small (smaller than 0.05). In the C. elegans neural network, connection between communities is more necessary and frequent due to some special functions.

\section{Applications in Weighted networks}

Our method can be generalized to weighted networks because the adjacency matrix in an undirected weighted network is real and symmetric. Thus, in weighted networks, the importance of a node and its role in communities are also characterized by its $I$ and $w$-score. Let us first consider an artificial weighted network. We use similarity weight in this weighted network. A higher weight means a closer relationship between vertices. At first, 10 nodes form a complete network and are divided into two communities with 5 nodes each. We assign vertices 4 and 9 as the core of each community, each of which has links with weight 2 connecting to vertices within its community and weight 0.2 connecting to outside vertices. All other intra-connections have weight 1, and all other interconnections have weight 0.2. Then we introduce vertex 11 as the bridge between the two communities. It connects to all 10 nodes with weight 1. The index $I$ and $w$-score for each node are given in Tab. 2. The results indicate that vertices 4, 9 and 11 are more important than the other vertices, while vertex 11 is a ``bridge" between these two communities. Our method works quite well in this small artificial weighted network.

\begin{table}[htbp]
\begin{center}
\caption{Centrality metrics $I$ and $w$-score in a complete weighted network.}
\begin{tabular}{p{2.5cm} p{1cm} p{1cm} p{2cm}}
\hline
\hline
Vertex Label & I & & $w$-score  \\
\hline
4 & 0.295 & &0.316\\
9 & 0.295 & &0.316\\
11& 0.16 & &0.00\\
others & 0.156& &0.316\\
\hline
\hline
\end{tabular}
\vspace*{0.0cm}
\end{center}
\end{table}

\begin{figure}
\center\includegraphics[width=0.5\textwidth]{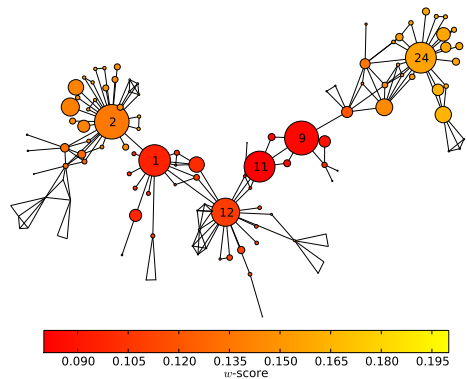}
\caption{Sketch of the SFI scientific collaboration network as a weighted, undirected network. It has 118 scientists. Vertex diameters indicate the community centrality $I$. The color of each vertex is proportional to the index $w$-score.}
\end{figure}

As an example of a real-world weighted network, we investigate the collaboration network among scientists working at the Santa Fe Institute (the SFI network). Here we consider it as a weighted, undirected network. Collaboration events between the scientists can be repeated again and again, and a higher frequency of collaboration usually indicates a closer relationship. Furthermore, weights can be assigned to the scientists' collaboration quite naturally: an article with $n$ authors corresponds to a collaboration act of weight $\frac{1}{{n - 1}}$ between every pair of its authors\cite{PRE73}. The results for the SFI collaboration network are sketched in Fig. 7. Vertex diameters indicate the community centrality $I$. The color of each vertex is proportional to the index w-score. Red vertices behave like ``overlapping" nodes or ``bridges" between communities, and yellow vertices often lie inside their own communities. We do not know the specific names; however, we observe that the positions of the large vertices are just like the ``group leaders". Vertices 2, 12 and 24 are so-called ``community cores" in communities because their $w$-scores are quite large. In fact, they are the group leaders in the fields of Mathematical Ecology, Statistical Physics and Structure of RNA, respectively. However, vertices 1, 9 and 11 are the ``bridges" between communities, and they have relative small $w$-scores. Interestingly, the result in the weighted network is different from the one in the corresponding unweighted network. It can be concluded that the edge weight may affect the result. For example, vertex 9 and vertex 11 collaborate quite often; this makes both of them quite important in a weighted network, while in an unweighted network, neither of them is very important to the community structure.

\section {Conclusion And Discussion}

In this paper, we characterize the node importance to community structure using the spectrum of the graph. The eigenspectrum of the adjacency matrix gives a clear indication of the number of ``dominant" communities in a network\cite{PRE80}. We give a centrality metric based on the spectrum of the adjacency matrix of the graph, and it can identify the nodes important to the community structure in many cases. In addition, we propose an index to distinguish the two kinds of important nodes that we term ``community cores" and ``bridges" using the spectrum of the graph Laplacian.

We demonstrate a variety of applications of our method to both artificial and real-world networks representing social and neural networks. Our method works well in many cases without knowing the exact community structure, although the number of communities should be known. However, a limitation of this method arises when one or more of the communities is much smaller than the largest community, or when a community has very sparse intra-community connections compared to other communities. This may happen when $N_{small}^2 < N_{large}$\cite{PRE80}. Even in the absence of perturbation, the maximum eigenvalue of a smaller community can lie inside the cloud of non-Perron-Frobenius eigenvalues of the largest community. But, with the understanding that the intent of our method is to find the important nodes in the community structure, the nodes in very small communities may be ignored. Even so, if the community structure is so fuzzy that we cannot identify the number of communities, our method is not accurate.

Our method can also be used in weighted networks. From our result in the SFI network, it can be inferred that edge weight may affect the result. Furthermore, it may generalize to directed networks because the Perron-Frobenius eigenvalues are often real and positive\cite{SIAMR}. We have yet to treat the case of directed networks. The identification of such key nodes is important and could potentially be used to identify the organizer of the community in social networks, to develop an immunization strategy in an epidemic process, to identify key nodes in biological networks and so on. We hope our results may be helpful to future research.
\\
\\
\section*{ACKNOWLEDGEMENTS}
The authors thank Di Huan, An Zeng, and Hongzhi You for their
helpful suggestions. This work is supported by the NSFC under grants No. 70771011 and No. 60974084, NCET-09-0228, and fundamental research funds for the Central Universities of Beijing Normal University.


\begin{thebibliography}{99}

\bibitem{Rev.Mod.Phys.74}Albert R, Barabasi A-L (2002), Statistical mechanics of complex networks. Rev. Mod.  Phys.\textbf{74}: 47-97.
\bibitem{SIAM Rev.45}Newman MEJ, The structure and function of complex networks (2003). SIAM Rev. \textbf{45}: 167-256.
\bibitem{Science 286}Barabasi A-L, Albert R (1999), Emergence of Scaling in Random Networks. Science \textbf{286}: 509-512.
\bibitem{Nature 393}Watts DJ, Strogatz SH (1998), Collective dynamics of `small-world' networks. Nature \textbf{393}: 440-442.
\bibitem{Phys. Rep. 424}Boccaletti S, Latora V, Moreno Y, Chavez M and Hwang D-U (2006), Complex networks: Structure and dynamics. Physics Reports \textbf{424}: 175-308.
\bibitem{PNAS99}Girvan M, Newman MEJ (2002), Community structure in social and biological networks. Proc. Natl. Acad. Sci. \textbf{99}: 7821-7826.
\bibitem{Plos1}Lancichinetti A, Kivela M, Saramaki J, Fortunato S (2010), Characterizing the Community Structure of Complex Networks. PloS ONE, \textbf{5}: e11976.
\bibitem{Phys. Rev. E 74}Newman MEJ (2006), Finding community structure in networks using the eigenvectors of matrices. Phys. Rev. E \textbf{74}: 036104.
\bibitem{Physics Reports486}Fortunato S (2009), Community detection in graphs. Physics Reports \textbf{486}: 75-174.
\bibitem{Eur.Phys.J.B.38}Wu F and Huberman BA (2004), Finding communities in linear time: A physics approach. Eur.Phys.J.B.\textbf{38}: 331-338.
\bibitem{Phys.Rev.E 72}Duch J and Arenas A (2005), Community detection in complex networks using extremal optimization. Phys. Rev. E \textbf{72}: 027104.
\bibitem{Proc. Natl. Acad. 103}Newman MEJ (2006), Modularity and community structure in networks. Proc. Natl. Acad. \textbf{103}: 8577-8582.
\bibitem{Phys. Rev. E 72}Gfeller D, Ghappelier J-C and Los Rios P de (2005), Finding instabilities in the community structure of complex networks. Phys. Rev. E \textbf{72}: 056135.
\bibitem{arXiv:1002.2007v1}Hu Y, Ding Y, Fan Y and Di Z (2010), How to Measure Significance of Community Structure in Complex Networks. arXiv:1002.2007v1.
\bibitem{arXiv:0902.3331v1}Hu Y, Nie Y, Yang H, Cheng J, Fan Y and Di Z (2010), Measuring the significance of community structure in complex networks. Phys. Rev. E \textbf{82}: 066106.
\bibitem{Rhys. Rev. E 77}Karrer B, Levina E and Newman MEJ (2008), Robustness of community structure in networks. Rhys. Rev. E \textbf{77}: 046119.
\bibitem{arXiv:0907.3708}Lancichinetti A, Radicchi F, Ramasco JJ (2010), Statistical significance of communities in networks. Phys. Rev. E \textbf{81}: 046110.
\bibitem{Proc. Natl. Acad. Sci.100}Spirin V, Mirny LA (2003), Protein complexes and functional modules in molecular networks. Proc. Natl. Acad. Sci.\textbf(100); 12123-12128.
\bibitem{Physica A 384}Sun L, Li M, Jiang L, Tan L (2007), Comparative analysis of the gene co-regulatory network of normal and cancerous lung. Physica A \textbf{384}: 739-746.
\bibitem{Europhys. Lett.72}Liu Z and Hu B (2005), Epidemic spreading in community networks. Europhys. Lett.\textbf{72}: 315.
\bibitem{Nature433}Guimera R, Amaral LAN (2005), Functional cartography of complex metabolic networks. Nature \textbf{433}: 895-900.
\bibitem{PLOSONE}Kovacs IA, Palotai R, Szalay MS, Csermely P (2010), Community landscapes: an integrative approach to determine overlapping network module hierarchy, identify key nodes and predict network dynamics. PLoS ONE \textbf{5}: e12528.
\bibitem{Nature430}Han JD, Bertin N, Hao T, Goldberg DS, Berriz GF (2004), et al. Evidence for dynamically organized modularity in the yeast protein¨Cprotein interaction network. Nature, \textbf{430}: 88-93.
\bibitem{PLoS Bio}Batada NN, Reguly T, Breitkreutz A, Boucher L, Breitkreutz BJ (2006), et al. Stratus not altocumulus: A new view of the yeast protein interaction network. PLoS Biol \textbf{4}: e317.
\bibitem{PLoS Bioo}Batada NN, Reguly T, Breitkreutz A, Boucher L, Breitkreutz BJ, Hurst LD, Tyers M (2007), Still stratus not altocumulus: further evidence against the date/party hub distinction. PLoS Biol \textbf{5}: e154.
\bibitem{PRE82}Zhao M, Zhou C, Chen Y, Hu B, Wang B (2010), Complexity versus modularity and heterogeneity in oscillatory networks: Combining segregation and integration in neural systems. Physical Review E \textbf{82}: 046225.
\bibitem{PRE80}Chauhan S, Girvan M, Ott E (2009), Spectral properties of networks with community structure. Physical Review E, \textbf{80}: 056114.
\bibitem{Neuro}Stam CJ (2005), Nonlinear dynamical analysis of EEG and MEG: Review of an emerging field. Clin. Neurophysiol. \textbf{116}: 2266-2301.
\bibitem{CMJ23}Fiedler M (1973), Algebraic connectivity of graphs. Czech. Math. J. \textbf{23}: 298-305.
\bibitem{Ratiocut}Hagen L, Kahng A (1992), New spectral methods for ratio cut partitioning and clustering. IEEE Trans. Computer-Aided Design, \textbf{11}: 1074-1085.
\bibitem{Tutorial}Luxburg UV (2007), A Tutorial on spectral clustering. Statistics and Computing, \textbf{17}: 395-416.
\bibitem{Phys. Rev. E 78}Lancichinetti A, Fortunato F and Radicchi F (2008), Benchmark graphs for testing community detection algorithms. Phys. Rev. E \textbf{78}: 046110.
\bibitem{JAR33}Zachary WW (1977), An information flow model for conflict and fission in small groups. Journal of Anthropological Research \textbf{33}: 452-473.
\bibitem{SOUTH}Nelson DL, McEvoy CL, Schreiber TA (1998), The university of south florida word association, rhyme, and word fragment norms.
\bibitem{website}http://www-personal.umich.edu/~mejn/netdata/
\bibitem{TRS}White JG et al. (1986), The structure of the nervous system of the nematode caenorhabditis elegans. Philos. Trans. R. Soc. London, Ser. B \textbf{314}: 1-340.
\bibitem{JN}Tsalik EL and Hobert OL, Neurobiol J (2003). Functional Mapping of Neurons That Control Locomotory Behavior in Caenorhabditis elegans. \textbf{56}: 178-197.
\bibitem{PRE73}Ramasco JJ and Morris SA (2006), Social inertia in collaboration networks. Phys. Rev. E \textbf{73}: 016122.
\bibitem{SIAMR}MacCluer CR (2000), The many proofs and applications of Perron's Theorem. SIAM Rev. \textbf{42}: 487-498.

\end{thebibliography}
\end{document}